\documentclass[aps,prl,twocolumn,showpacs,
preprintnumbers,amsmath,amssymb]{revtex4-1}
\usepackage[colorlinks=true, pdfstartview=FitV, linkcolor=red, citecolor=blue, urlcolor=blue, pdftitle={A No-Go Theorem for Critical Phenomena in Large-$N_c$ QCD},pdfauthor={Yoshimasa Hidaka, Naoki Yamamoto},pdfsubject={}, pdfkeywords={Large N QCD, chiral phase transition, finite temperature and density, orbifold equivalence}]{hyperref}

\usepackage{amsmath,amssymb,bm}
\usepackage[dvips]{graphicx}
\usepackage{yfonts}
\usepackage{enumerate}
\newcommand{\beq}{\begin{eqnarray}}
\newcommand{\eeq}{\end{eqnarray}}

\newcommand{\tr}{\mathop{\mathrm{tr}}}
\newcommand{\SU}{\text{SU}}
\newcommand{\SO}{\text{SO}}
\newcommand{\Sp}{\text{Sp}}
\newcommand{\U}{\text{U}}

\newcommand{\arXiv}[2]{\href{http://arxiv.org/abs/#1}{#2}}
\begin{document}

\preprint{RIKEN-MP-35, INT-PUB-11-047}
\author{Yoshimasa Hidaka$^{1}$ and Naoki Yamamoto$^{2}$}
\affiliation{
$^{1}$Mathematical physics Laboratory, RIKEN Nishina Center, Saitama 351-0198, Japan\\
$^{2}$Institute for Nuclear Theory, University of Washington, 
Seattle, Washington 98195-1550, USA}
\date{March 19, 2012}

\title{No-Go Theorem for Critical Phenomena in Large-$N_c$ QCD}
\begin{abstract}
We derive some rigorous results on the chiral phase transition in 
QCD and QCD-like theories with a large number of colors, $N_c$, based on 
the QCD inequalities and the large-$N_c$ orbifold equivalence. 
We show that critical phenomena and associated soft modes are forbidden 
in flavor-symmetric QCD at finite temperature $T$ and finite but not 
so large quark chemical potential $\mu$ for any nonzero quark mass.
In particular, the critical point in QCD at a finite baryon chemical 
potential $\mu_B = N_c \mu$ is ruled out, if the coordinate ($T, \mu$)
is outside the pion condensed phase in the corresponding phase diagram 
of QCD at a finite isospin chemical potential $\mu_I = 2\mu$.

\end{abstract}
\pacs{11.10.Wx, 11.15.Pg, 12.38.Lg}
\maketitle

\emph{Introduction.}---%
The phase structure of quantum chromodynamics (QCD) 
at finite temperature $T$ and finite baryon chemical potential $\mu_B$ 
is a longstanding problem despite its phenomenological importance,
including heavy ion collisions and cosmology.
Although it has been established from the first-principles lattice 
QCD simulations that the thermal chiral transition at $\mu_B=0$ 
is a smooth crossover in real QCD \cite{Aoki:2006we}, the fate 
of the chiral transition at nonzero $\mu_B$ has not been fully understood. 
In particular, not only the location, but even the existence of 
the QCD critical point(s) (see \cite{Stephanov:2004wx} for a review)
has not yet been settled. This is mainly because the Monte Carlo 
method is not available at nonzero $\mu_B$ due to the sign problem.

One might hope that the $1/N_c$ expansion provides some new insights 
to this question, where the number of colors $N_c$ is taken to 
infinity with keeping the 't Hooft coupling $\lambda = g^2 N_c$ fixed 
(the 't Hooft limit) \cite{'tHooft:1973jz}.
This limit has proven successful for understanding of a number of 
aspects of hadrons in the QCD vacuum \cite{Witten:1979kh},
and has also been widely applied to QCD at finite $T$ 
and finite baryon chemical potential $\mu_B = N_c \mu$.
The deconfinement temperature $T_d$ is independent of $\mu$ 
(when $\mu \sim N_c^0$) in this limit
since the gauge dynamics with $\sim N_c^2$ degrees of freedom 
is insensitive to the quark dynamics with $\sim N_c^1$ 
degrees of freedom \cite{McLerran:2007qj}. 
However, the fate of the chiral phase transition is still 
an unanswered question. We only know that the critical temperature 
$T_c$ of the chiral transition must satisfy $T_c \geq T_d$ because 
the chiral condensate $\sim N_c^1$ cannot be changed by noninteracting 
mesons and glueballs with $\sim N_c^0$ degrees of freedom 
in the confined phase \cite{Neri:1983ic}.

In this Letter, we derive some exact results on the chiral phase 
transition in the large-$N_c$ QCD and QCD-like theories.
The critical phenomena (especially the QCD critical point) 
and associated soft modes are forbidden in flavor-symmetric QCD 
at finite $\mu_B$ for any nonzero quark mass $m$, as long as 
the coordinate ($T, \mu$) is outside the pion condensed phase 
\cite{note:N_c} in the corresponding phase diagram of QCD at finite 
isospin chemical potential $\mu_I = 2\mu$
(for the phase diagram, see Fig.~\ref{fig:phase} below).

\begin{figure}[b]
\begin{center}
\includegraphics[width=7.5cm]{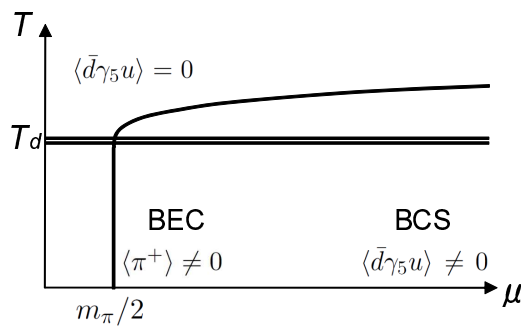}
\end{center}
\vspace{-0.5cm}
\caption{Phase diagram of QCD with $\mu_I$ at large $N_c$.
The single line denotes the pion condensed phase transition 
and the double line denotes deconfinement transition. 
(The chiral transition is not shown here.)}
\label{fig:phase}
\end{figure}

Our results still allow a possibility that the QCD critical point
exists inside the pion condensed phase in the large-$N_c$ limit.
Actually, in various effective models, 
such as the random matrix model \cite{Han:2008xj}, 
Nambu--Jona-Lasinio (NJL) model \cite{Andersen:2009zm},
and PNJL model \cite{Sakai:2010kx}, the critical point has been 
observed inside this region, as is consistent with our no-go theorem. 
This can be ascribed to the fact that the mean-field approximation in 
the model calculations corresponds to the leading order of $1/N$ expansion,
and no-go theorems can also be formulated within these models \cite{HY}.
Since the sign problem is maximally severe inside the
pion condensed phase as pointed out by model analyses 
\cite{Han:2008xj, Andersen:2009zm, Sakai:2010kx}, our no-go theorem 
might imply that the conventional reweighting techniques 
are difficult to access the QCD critical point on the lattice.
Our rigorous results are also useful to judge whether holographic 
models of QCD \cite{Aharony:2006da} motivated by the gauge/gravity 
duality \cite{Maldacena:1997re} capture the genuine QCD physics or not.

\emph{QCD inequalities.}---%
We first recall the rigorous QCD inequalities \cite{Weingarten:1983uj} 
which are essential in the discussion of this Letter. 
We shall work in the Euclidean and flavor-symmetric QCD 
with $N_f$ flavors, and consider the Dirac operator at $\mu_B=0$, 
${\cal D} = D + m$ with 
$D=\gamma_{\mu} (\partial_{\mu}+ig A_{\mu})$. 
The operator $D$ satisfies the 
anti-Hermiticity and chiral symmetry, 
$D^{\dag}=-D$ and $\gamma_5 D \gamma_5 = -D$.
From these two properties, we have
\beq
\label{eq:gamma5}
\gamma_5 {\cal D} \gamma_5 = {\cal D}^{\dag},
\eeq
and the positivity, $\det {\cal D} \geq 0$.

Let us take a generic flavor nonsinglet fermion bilinear 
$M_{\Gamma}=\bar \psi \Gamma \psi$ and
consider a set of correlation functions,
\beq
C_{\Gamma}(x, y) &\equiv& \langle M_{\Gamma}(x) M_{\Gamma}^{\dag}(y) \rangle_{\psi, A}\notag\\
&=& - \langle \tr [S_A(x,y) \Gamma S_A(y,x) \bar \Gamma] \rangle_A.
\eeq
Here, $S_A(x,y) \equiv \langle x|{\cal D}^{-1} |y \rangle$ 
is a propagator from $y$ to $x$ in a background gauge field $A$,
the symbols $\langle \ \cdot \ \rangle_{\psi, A}$ and $\langle \ \cdot \ \rangle_{A}$
denote the full average and the average over the gauge field, respectively,
and $\bar \Gamma \equiv \gamma_0 \Gamma^{\dag} \gamma_0$.
From (\ref{eq:gamma5}) and the positivity of the measure, 
we have
\beq
\label{eq:inequality}
C_{\Gamma} &=& \langle \tr [S_A(x,y) \Gamma i\gamma_5 
S^{\dag}_A(x,y) i\gamma_5 \bar \Gamma] \rangle_A \nonumber \\
&\leq &
\langle \tr [S_A(x,y) S^{\dag}_A(x,y)] \rangle_A,
\eeq
where the Cauchy-Schwarz inequality is used.
The inequality is saturated when $\Gamma = i\gamma_5 \tau_A$ 
with $\tau_A$ being the traceless flavor generators.

The asymptotic behavior of $C_{\Gamma}$ at large distance $|x-y|$ can be written as
\beq
C_{\Gamma} \sim e^{-m_{\Gamma}|x-y|},
\eeq
with $m_{\Gamma}$ being the mass of the lowest meson state in the channel $\Gamma$.
Then the inequalities among correlators (\ref{eq:inequality}) 
lead to the inequalities among meson masses,
\beq
m_{\Gamma} \geq m_{\pi},
\eeq
where $m_{\pi}$ is the mass of the pseudoscalar pion.

We note that the derivation of the QCD inequalities so far
relies on the assumption that $j_{\Gamma}$ is not flavor singlet.
This condition is necessary, otherwise the flavor disconnected diagrams
$\sim \langle \tr[\Gamma S_A(x,x)]\tr[\bar \Gamma S_A(y,y)] \rangle_A$, 
where $\bar \psi \psi$ turns into a gluonic intermediate state, 
also contribute and the above argument breaks down.
Phenomenologically, the disconnected diagrams might be 
suppressed compared with the connected diagrams. 
A well-known example is the Okubo-Zweig-Iizuka (OZI) rule. 
Theoretically, it is the 't Hooft large-$N_c$ limit \cite{'tHooft:1973jz} that 
cleanly justifies this statement \cite{Witten:1979kh}.
The flavor disconnected diagrams are subleading compared with the connected 
diagrams in the $1/N_c$ expansion.
Hence, at the leading order in the $1/N_c$ expansion, the QCD inequalities are 
also applicable to the flavor singlet channel. In particular, it follows that
\beq
\label{eq:bound1}
m_{\sigma} \geq m_{\pi}.
\eeq
Here, $m_{\sigma}$ is the mass of the flavor singlet scalar $\sigma$
which has the quantum number of the chiral condensate $\langle \bar \psi \psi \rangle$.
If one includes the subleading $1/N_c$ corrections that originate 
from the flavor disconnected diagrams, $m_{\sigma}$ can be written as
\beq
\label{eq:sigma-general}
m_{\sigma} = m_{\pi} + C + {\cal O}({N_c^{-1}}),
\eeq
with some $C \geq 0$. We here assume $m={\cal O}(N_c^0)$, 
and thus $m_{\pi}={\cal O}(N_c^0)$,
since quark mass does not have any dependence of $N_c$ in nature.

\emph{Chiral phase transition at $\mu_B=0$.}---%
Let us consider the thermal chiral phase transition in large-$N_c$
QCD at $\mu_B=0$ in the presence of nonzero quark mass $m$.
In this case, a pion becomes massive due to the explicit breaking of 
chiral symmetry, $m_{\pi} > 0$.
If there exists a second-order chiral transition 
at some critical temperature $T=T_c$ at $\mu_B=0$,
the screening mass in the $\sigma$-meson channel vanishes, $m_{\sigma} =0$,
or the correlation length $\xi=m_{\sigma}^{-1}$ is divergent 
\cite{Rajagopal:1992qz}.

However, the inequality (\ref{eq:bound1}), which is valid independent of $T$,
leads to the finite bound $\sim N_c^0$ for $m_{\sigma}$,
\beq
\label{eq:bound2}
m_{\sigma} \geq m_{\pi} >0.
\eeq
This constraint clearly contradicts the fact that $m_{\sigma}=0$ at $T=T_c$.
Therefore, we arrive at the conclusion that the second-order chiral 
transition and associated soft modes are forbidden in the large-$N_c$ QCD 
at $\mu_B=0$ for any $m > 0$. This is our first no-go theorem.
From this theorem, the large-$N_c$ thermal chiral transition
at $\mu_B=0$ for $m>0$ is either first order or crossover;
we will discuss each possibility later.

Note that we used the large-$N_c$ limit only to justify 
the suppression of flavor disconnected diagrams compared with flavor connected ones.
This implies that, in real QCD with $N_c=3$, the second-order chiral transition
for $m>0$ can happen, when the contribution of disconnected 
diagrams, the ${\cal O}({N_c}^{-1})$ term in (\ref{eq:sigma-general}), 
cancels out that of connected diagrams, the remaining terms in (\ref{eq:sigma-general}). 

Our no-go theorem agrees with the general argument on the order of 
the chiral phase transition based on the symmetries of QCD \cite{Pisarski:1983ms}. 
In three-color and three-flavor QCD, the thermal chiral transition at $\mu_B=0$ 
becomes first order because of the $\U(1)_A$ anomaly for small $m$, 
and becomes second order at some critical quark mass $m_c$. 
At large $N_c$, the anomaly effects related to disconnected diagrams are suppressed 
as ${\cal O}(N_c^{-1})$, and thus, $m_c = \mathcal{O}(N_c^{-1})$; 
the chiral transition is smeared into a crossover 
for any nonzero quark mass $\sim N_c^{0}$.

\emph{Chiral phase transition at $\mu_I \neq 0$.}---%
The positivity of the theory, which enables us to use the QCD
inequalities above, is essential to derive the no-go theorem for the
chiral critical phenomena at $\mu_B=0$.
However, the Dirac operator at finite $\mu_B=N_c \mu$, 
${\cal D}(\mu) = D + \mu \gamma_0 + m$
does no longer satisfy (\ref{eq:gamma5}),
and the positivity is lost; 
this is the origin of the notorious sign problem in QCD. 

On the other hand, there are class of QCD-like theories that have 
the positivity at finite $\mu_B$, 
such as QCD with fermions in the adjoint representation \cite{Kogut:2000ek}, 
$\SO(2N_c)$ \cite{Cherman:2010jj, Hanada:2011ju}
and $\Sp(2N_c)$ gauge theories \cite{Hanada:2011ju}. 
Also QCD at finite isospin chemical potential $\mu_I$ maintains
the positivity \cite{Son:2000xc}.
(We do not consider two-color QCD at finite $\mu_B$, which also 
has the positivity \cite{Kogut:1999iv, Kogut:2000ek},
because we cannot take the large-$N_c$ limit in this theory.)
We can apply our previous argument 
to these theories even at finite $\mu_B$ or finite $\mu_I$,
except the adjoint QCD. The reason why our argument fails in 
the adjoint QCD is that the flavor disconnected diagrams are 
not suppressed compared with the connected diagrams since ``color" degrees of 
freedom of fermions $\sim N_c^2$ are comparable to those of gluons.
Similarly, our argument is not applicable to QCD with fundamental quarks 
for fixed $N_f/N_c$ and $N_c \rightarrow \infty$ \cite{Veneziano:1976wm}
and QCD with two-index antisymmetric quarks for fixed $N_f$ 
and $N_c \rightarrow \infty$ \cite{Corrigan:1979xf},
the latter of which is used for the orientifold equivalence
with adjoint QCD \cite{Armoni:2003fb}.

Let us take QCD at finite $\mu_I$ with two degenerate flavors  
as an example, which will be utilized later to generalize the no-go theorem to 
QCD at {\it finite} $\mu_B$. 
The same argument is applicable to other theories,
$\SO(2N_c)$ gauge theory with any number of flavors and
$\Sp(2N_c)$ gauge theory with even number of flavors \cite{note:positivity}.

The Dirac operator in QCD at finite $\mu_I$, 
${\cal D}(\mu_I) = D + \mu_I \gamma_0 \tau_3/2 + m$
satisfies the relation \cite{Son:2000xc}
\beq
\label{eq:isospin}
\tau_1 \gamma_5 {\cal D} \gamma_5 \tau_1 = {\cal D}^{\dag},
\eeq 
for the degenerate quark mass $m$, from which 
the positivity $\det{\cal D}(\mu_I) \geq 0$ follows. 
One can then derive the inequality at {\it any} $\mu_I$ \cite{Son:2000xc}:
\beq
C_{\Gamma} \leq \langle \tr [S_A(x,y) S^{\dag}_A(x,y)] \rangle_A,
\eeq
where the inequality is saturated when $\Gamma=i\gamma_5 \tau_{1,2}$.
This leads to the inequalities between meson masses in different channels,
$m_{\Gamma} \geq m_{\pi_{\pm}}$. 
In the large $N_c$ limit, QCD inequalities are 
applicable to the flavor singlet channel, and we have
\beq
m_{\sigma} \geq m_{\pi_{\pm}}.
\label{eq:inequality_pi}
\eeq
Repeating a similar argument to QCD at $\mu_B=0$, 
the second-order chiral transition is not allowed in this theory 
at finite $\mu_I$ where $m_{\pi_{+}} > 0$ or $m_{\pi_{-}} > 0$.

In order to consider the applicable region 
of the above no-go theorem, let us turn to the phase diagram of 
QCD at finite $\mu_I$ in the large-$N_c$ limit shown 
in Fig.~\ref{fig:phase}. (See \cite{Son:2000xc} for 
the phase diagram with $N_c=3$.)
In this case, $\pi_+$ meson charged under isospin symmetry 
exhibits the Bose-Einstein condensation (BEC) $\langle \pi_+ \rangle \neq 0$ 
at low density, $\mu>m_{\pi}/2$ at $T=0$ where the excitation
energy $m_{\pi} - 2\mu$ becomes negative. On the other hand, 
at high density, the attractive interaction between quarks near 
the Fermi surface leads to the Bardeen-Cooper-Schrieffer (BCS) 
pairing of diquark with the quantum number 
$\langle \bar d \gamma_5 u \rangle$, as found from the 
weak-coupling calculations \cite{Son:2000xc}. 
Because both condensates have the same quantum numbers and break
the same symmetry $\U(1)_{L+R}$ down to ${\mathbb Z}_2$, there should be
no phase transition between the two regimes, similarly to the
BEC-BCS crossover studied in nonrelativistic Fermi gases \cite{Giorgini:2008zz}.

In Fig.~\ref{fig:phase}, the deconfinement temperature $T_d$ 
is independent of $\mu$, as explained in the introduction.
Also the critical chemical potential $\mu_c=m_{\pi}/2$
for the pion condensation is independent of $T$ in the confined phase
because of the large-$N_c$ volume independence \cite{Kovtun:2007py}
or from the argument similar to the chiral condensate \cite{Neri:1983ic}.

From the phase diagram of QCD at finite $\mu_I$, 
$m_{\pi_{\pm}}>0$ ($m_{\pi_+}=0$ and $m_{\pi_-}>0$) 
outside (inside) the pion condensed phase.
Therefore, from the QCD inequalities (\ref{eq:inequality_pi}), 
massless $\sigma$ and the second-order chiral transition 
are prohibited in QCD at any $\mu_I$ for any $m>0$.

Repeating the same argument in $\SO(2N_c)$ and $\Sp(2N_c)$ gauge theories 
at finite $\mu_B$, the second-order chiral transition is forbidden in these
theories. The locations of phase boundaries of the chiral transition, 
deconfinement transition, and BEC-BCS crossover region completely 
coincide with those of QCD at finite $\mu_I$, 
as shown in \cite{Hanada:2011ju}.

\emph{Chiral phase transition at $\mu_B \neq 0$.}---%
Now we are ready to generalize the no-go theorem to QCD at finite $\mu_B$, 
using the results of QCD at finite $\mu_I$.
Based on the large-$N_c$ orbifold equivalence \cite{Bershadsky:1998cb},
it was recently shown that a class of observables in QCD at finite $\mu_B$, 
including the chiral condensate, exactly coincide those of QCD at finite $\mu_I$ 
outside the pion condensed phase.
This can also be understood from the following argument: At the leading order of 
$1/N_c$, the contributions of up (u) and down (d) quarks to the chiral condensate are {\it decoupled}
from each other, and hence, the chiral condensate does not distinguish the sign of the 
chemical potential for the d quark due to the charge conjugation symmetry. 
As a consequence, chiral condensate at finite $\mu_B$ and that at finite $\mu_I$ coincide.
However, this argument fails if the pion condensation occurs in QCD at finite $\mu_I$
where u and d quarks are {\it coupled} in the ground state.

Since we already know from the argument above that chiral critical phenomena 
are forbidden in the large-$N_c$ QCD at finite $\mu_I$ for $m>0$, 
the same must be true in QCD at finite $\mu_B$ outside the 
pion condensed phase in the corresponding phase diagram of QCD at finite $\mu_I$.
Similarly one can obtain the no-go theorem in QCD at finite $\mu_B$
from that of $\SO(2N_c)$ [or $\Sp(2N_c)$] gauge theory at finite $\mu_B$
by using the large-$N_c$ orbifold equivalence \cite{Cherman:2010jj, Hanada:2011ju}
between these theories outside the diquark condensed phase. 
In particular, the equivalence with $\SO(2N_c)$ gauge theory
at finite $\mu_B$ leads to the stronger no-go theorem that, 
chiral critical phenomena are not allowed in massive and flavor-symmetric 
QCD with {\it any} number of flavors at finite $\mu_B$, if the coordinate 
($T, \mu$) is outside the diquark condensed phase in the 
corresponding phase diagram of $\SO(2N_c)$ gauge theory.

\emph{Chiral phase transition in the chiral limit.}---%
Let us turn to the chiral transition in the chiral limit $m=0$.
We first recall that chiral transition is either first or 
second order at $m=0$. When $T_c > T_d$, we can utilize
our no-go theorem for $m>0$ to constrain the chiral transition
at $m=0$; a first-order chiral transition at $m=0$ is prohibited 
by the no-go theorem because, with increasing $m$, it would 
eventually become second order at some critical $m=m_c$,
while a second-order chiral transition at $m=0$ is allowed because 
it would be smeared into a crossover for any nonzero 
$m={\cal O}(N_c^0)$ \cite{note:mass}.
When $T_c=T_d$, however, an interplay between chiral symmetry breaking 
and deconfinement always makes the chiral transition first order for 
any $m$ \cite{HY} and is not constrained by the no-go theorem.

Therefore, we arrive at two possible scenarios for 
the chiral transition at $m=0$:
(a) the chiral transition is second order when $T_c>T_d$, or
(b) the chiral transition is first order when $T_c=T_d$.
Which scenario is realized in the large-$N_c$ QCD cannot be determined 
from our arguments alone, and should be studied from other constraints 
or in the numerical lattice QCD simulations.

\acknowledgments
We thank A.~Cherman, M.~Hanada, K.~Hashimoto, H.~Iida, D.~B.~Kaplan, 
A.~Karch, T.~Morita, and D.~T.~Son for useful discussions, 
and T.~Brauner and R.~D.~Pisarski for comments. 
N.Y. thanks the hospitality of Mathematical Physics Group
at RIKEN Nishina Center where this work was initiated. 
Y.H. is  supported by a Grant-in-Aid for Scientific Research by
the Ministry of Education, Culture, Sports, Science and Technology (MEXT) of
Japan (No.23340067).
N.Y. is supported by JSPS Postdoctoral Fellowships for Research Abroad.


\begin{thebibliography}{99}
\bibitem{note:mass}
Here we assume $m={\cal O}(N_c^0)$ and do not consider the region 
$m={\cal O}(N_c^{-1})$. The latter is different from the usual 't 
Hooft large-$N_c$ counting so that one has to reorganize the 
diagrammatic $1/N_c$ expansion itself; e.g., the large-$N_c$ 
orbifold equivalence could be spoiled in this region.
This is beyond the scope of the present Letter.
\end{thebibliography}
\end{document}